\documentclass[fleqn,usenatbib]{mnras}

\usepackage{newtxtext,newtxmath}

\usepackage[T1]{fontenc}

\DeclareRobustCommand{\VAN}[3]{#2}
\let\VANthebibliography\thebibliography
\def\thebibliography{\DeclareRobustCommand{\VAN}[3]{##3}\VANthebibliography}

\usepackage{graphicx}	
\usepackage{amsmath}	
\usepackage{xcolor}

\title[Dark subhaloes in clusters]{Tiny galaxies and dark substructures: exploring the ``dark'' subhaloes in TNG50}

\author[J. E. Doppel et al.]{
Jessica E. Doppel,$^{1,2}$\thanks{E-mail: jessicadoppel@gmail.com}
Mathilde Jauzac,$^{1,2,3,4}$ David J. Lagattuta,$^{1,2,5}$ Azadeh Fattahi,$^{2,6}$ and Guillaume Mahler$^{7,1,2}$
\\
$^{1}$Centre for Extragalactic Astronomy, Durham University, South Road, Durham DH1 3LE, UK\\
$^{2}$Institute for Computational Cosmology, Durham University, South Road, Durham DH1 3LE, UK\\
$^{3}$Astrophysics Research Centre, University of KwaZulu-Natal, Westville Campus, Durban 4041, South Africa \\
$^{4}$School of Mathematics, Statistics \& Computer Science, University of KwaZulu-Natal, Westville Campus, Durban 4041, South Africa \\
$^{5}$Centre for Astrophysics Research, Department of Physics, Astronomy and Mathematics, University of Hertfordshire, Hatfield AL10 9AB, UK\\
$^{6}$The Oskar Klein Centre, Department of Physics, Stockholm University, Albanova University Center, 106 91 Stockholm, Sweden\\
$^{7}$STAR Institute, Quartier Agora - All\'ee du six Ao\^ut, 19c B-4000 Li\`ege, Belgium\\
}

\date{Accepted XXX. Received YYY; in original form ZZZ}

\pubyear{\the\year{}}

\begin{document}
\label{firstpage}
\pagerange{\pageref{firstpage}--\pageref{lastpage}}
\maketitle

\begin{abstract}

Dark matter haloes and subhaloes that host no luminous counterpart are predicted within our current understanding of galaxy formation within a $\Lambda$CDM paradigm. Observational tests, such as gravitational lensing, have made potential detections of such objects around individual galaxies as well as in galaxy groups and clusters. The question of whether or not a dim counterpart might reside in these objects remains an open question. We investigate this point using the TNG50-1 simulation of the IllustrisTNG project. Under the assumption of TNG50's galaxy formation model, we do not find haloes or subhaloes above a total mass of $\rm 10^{9.7} \ M_{\odot}$ that are entirely dark. However, under realistic effective surface brightness cuts of $\rm \leq 29 \ mag \ arcsec^{-2}$, the inference of the most massive dark subhalo in galaxy groups and clusters becomes $\rm M_{DM} \gtrsim 2 \times 10^{10} \ M_{\odot}$.  Concentrating on galaxy groups and clusters, we find that dark subhaloes are ubiquitous with more massive dark subhaloes tending to preferentially reside further from the centers of clusters. We find that subhaloes in the mass range of $\rm 4.5 \times 10^{7} \leq M_{DM}/M_{\odot} \leq 2.1\times 10^{8}$ tend to be the most likely to reside in the strong lensing regions of galaxy groups and clusters, and argue that future dark subhalo searches should investigate this mass range.

\end{abstract}

\begin{keywords}
galaxies:clusters:general -- galaxies:dwarfs -- galaxies:halos
\end{keywords}



\section{Introduction}


$\Lambda$CDM \citep[][]{White1978, White1991} has proven one of the most successful cosmological frameworks to date \citep[e.g., ][]{Sales2022, GamezMartin2024, Sands2024}. It predicts the hierarchical formation of structure from tiny, low mass halos that merge into the galaxies, galaxy groups, and galaxy clusters that we see today. Adding model of galaxy formation onto the predictions of $\Lambda$CDM suggests that there should be dark matter haloes in which galaxies are unable to form \citep[e.g.,][]{Rees1986, Efstathiou1992, Thoul1996, Quinn1996, Barkana1999}, leading to a population of aptly named dark subhalos in semi-analytical and semi-empirical models \citep[e.g., ][]{Bullock2000, Benson2002, Nadler2020, Nadler2024} and cosmological hydrodynamical simulations \citep[e.g., ][]{Sawala2016, Jahn2019, ForouharMoreno2022}. Under different model assumptions (and resolutions) in these scenarios, the mass at which haloes cease to form galaxies varies \citep[e.g., ][]{Sawala2013,Sawala2016, Benitez-Llambay2020, Munshi2021,Kim2024}. Thus, it is an open question at what scale one should expect to see haloes that are entirely dark.

Testing the prediction for even the existence of dark-matter-only substructure in observations is, however, a difficult task. Indirect methods of detection are the only ways to find dark substructure. For example, around the Milky Way, the dynamics of stellar streams are thought to reveal the existence of tiny dark-matter-only objects \citep[e.g.,][]{Dehnen2004, Yoon2011, Carlberg2012, Bonaca2019}. In the less local universe, and particularly in massive galaxy clusters, distortions to gravitationally lensed images could indicate the presence of an interfering, sometimes dark, subhalos \citep{Vegetti2012, Vegetti2014, Despali2024, Lange2025}

The majority of lensing clusters tend to reside between $z \sim 0.1-6$ \citep[see e.g., ][]{Postman2012, Bleem2015, Coe2019, Sharon2020, Richard2021, Fox2022}. Only a handful of such clusters reside above $z\gtrsim 1$ \cite[][]{Gonzalez2015}. It is observationally challenging in this redshift range to detect low-mass galaxies. This adds a complicating factor to the detection of truly dark subhaloes: are they truly dark, or do they contain tiny galaxies? In the last decade (or so), there has been an influx of observations of Local Group dwarf and ultra-faint satellites \citep[see][and references therein]{Simon2019}. There is substantial overlap between the predicted dark matter masses of these tiny galaxies from abundance matching studies \citep[e.g.,][]{Wechsler2018, Shuntov2022, Monzon2024, Shuntov2025} and simulations \citep[e.g.,][]{Ahvazi2024, Kim2024, Rey2025}, and the masses of subhaloes possibly detected from lensing studies \citep[$\rm M_{sub} \sim 10^{9} M_{\odot}$, e.g., ][]{Vegetti2014}.

In order to understand the dwarf and dark (sub)halo populations in galaxy clusters and in the field, we need to turn to cosmological hydrodynamical simulations of large cosmological volumes. Ideally, investigating large enough volumes to contain massive systems like lensing clusters at $z \lesssim 1$ would provide the most direct comparison. However, simulations that fit this criteria often have particle masses $> 10^6 \rm M_{\odot}$ \citep[e.g.,][]{Schaye2015, Crain2015, pillepich2018, Schaye2023, Vogelsberger2014, Genel2014}; thus, low-mass dwarfs and dark subhaloes cannot be reliably resolved in these simulations. We therefore opt to use the IllustrisTNG50 simulation \citep{pillepich2019, nelson2019dr, nelson2019}, as it has a high enough resolution (m$_{DM} = 4.5 \times 10^5 \ \rm M_{\odot}$, $\rm m_{baryon} = 8.4\times 10^4 \ \rm M_{\odot}$)  to study the low-mass objects in question in realistic galaxy group and cluster environments.

This paper is organized as follows. In section \ref{sec:methods}, we briefly discuss the simulation and the selection criteria. In Sec. \ref{sec:results}, we discuss the expected mass function of dark subhaloes, the halo occupation fraction, and the distribution of dark substructures in galaxy clusters. We
present a brief discussion of the intepretation of our results in strong-lensing-based subhalo searches in Sec. \ref{ssec:lens}. Finally, we conclude with a summary of our results in Sec. \ref{sec:summary}

\section{Methods}
\label{sec:methods}
\subsection{Simulations}

We use the highest resolution run of the cosmological hydrodynamical simulation TNG50 \citep[][]{pillepich2019, nelson2019} due to its ability to simultaneously resolve low-mass objects and galaxy cluster environments. TNG50 is the smallest volume of the larger IllustrisTNG project \citep{naiman2018, pillepich2018, nelson2018, springel2018, marinacci2018, nelson2019dr}, featuring a box size of 51.7\,Mpc on one side and $2\times 2160^3$ gas cells and dark matter particles. The simulation is evolved from a flat, $\Lambda$CDM cosmology using cosmological parameters from \citet{planckcollab2016}. The simulation has a mass resolution of $4.5 \times 10^5 \ M_{\odot}$ for individual dark matter particles, an average mass of $8.4 \times 10^4 \ M_{\odot}$ for star and gas particles, and a gravitational softening length of $288$\,pc at $z = 0$.

Gravity and hydrodynamics in TNG50 is solved using AREPO. Heating and cooling of gas is followed down to $10^4$ K in an effective equation of state interstellar medium (ISM). TNG50's galaxy formation model includes star formation in the dense ISM, stellar evolution (including supernova and chemical enrichment of the ISM). Heating of the gas occurs via background radiation, and cooling includes metal line cooling and primordial cooling. Additionally, the model includes the seeding and growth of supermassive black holes, AGN feedback from both low and high accretion modes, and magnetic fields \citep{Pakmor2011, Pakmor2013, weinberger2017, pillepich2018b}.

\subsection{Galaxy and Subhalo Selection}

Haloes are identified in TNG50 using the Friends-of-Firends algorithm \cite[FoF, ][]{Davis1985} with a linking length of 0.2 times the mean interparticle separation, and subhalos within those haloes are further identified using \textsc{subfind} \citep{springel2001, Dolag2009}. Halo masses and radii are characterized by the total mass enclosed within a radius containing $200$ times the critical density of the universe.

In order to remain as close as possible to studies of dark substructures in galaxy clusters \citep[e.g., using strong gravitational lensing, see][]{Lagattuta2023, Mahler2023}, we chose to investigate the progenitors of the three most massive galaxy clusters at $z = 0$. The most massive cluster (M$_{\rm 200, c} = 1.83 \times 10^{14} \ \rm M_{\odot}$) is roughly consistent with low-end mass estimates of the Virgo Cluster \citep[M$_{200,c} \sim 2\times 10^{14} \ \rm M_{\odot}$,][]{Weinmann2011}, with the other two (M$_{\rm 200, c} = 9.41 \times 10^{14} \ \rm M_{\odot}$ and M$_{\rm 200,c} = 6.46 \times 10^{13} \ \rm M_{\odot}$) on par with massive galaxy groups, such as Fornax \citep[M$_{200,c} \sim 10^{14} \ \rm M_{\odot}$,][]{Drinkwater2001} at $z = 0$. We follow the main branch progenitors of these objects through in time using the SubLink merger tree \citep{rodriguez-gomez2015}. We consider the progenitors of these objects beginning at $z = 2$, and follow them at intervals of $dz \sim 0.2$. This allows us to understand how the population of low-mass subhalos associated to these structures changes over time.

Within the selected clusters, their progenitors, and the field, we choose (sub)haloes that have a total dark matter mass---the sum of all dark matter particles considered associated to an object via \textsc{subfind}---of at least $M_{\rm DM} = 4.54\times 10^7$ M$_{\odot}$, ensuring that they are resolved by at least 100 particles. We invoke an upper dark matter mass limit of $\rm M_{DM} = 2\times 10^{10} \ M_{\odot}$. The luminous components are considered resolved if they have 100 star particles associated to them, resulting in an average stellar mass of $M_{*} \gtrsim 8\times 10^6$ M$_{\odot}$.

We additionally consider an ``observationally'' resolved limit, which is made in addition to the aforementioned star particle number cut. We make this cut based on brightness limits \textit{as they are observed at $z = 0$}, which we take to be $\rm 29 \ mag \ arcsec^{-2}$. That is, we assume that half of the light (as recorded by the \textsc{subfind} catalogs) is contained within the stellar halfmass radius, $\rm r_{h,*}$, of a galaxy. We then take the average surface brightness within $\rm r_{h,*}$ and make a mock observation of the galaxy at its given redshift. For galaxies at $z = 0$, we assume they sit at the distance of the Virgo cluster \citep[16.5 Mpc, e.g.,][]{Mei2007}. We discuss the impact of this choice in Sections~\ref{sec:results} and \ref{sec:summary}.

\section{Results}
\label{sec:results}

\subsection{Dark Matter (Sub)halo Mass Function}
\label{ssec:mass_func}
\begin{figure}
    \centering
    \includegraphics[width = \columnwidth]{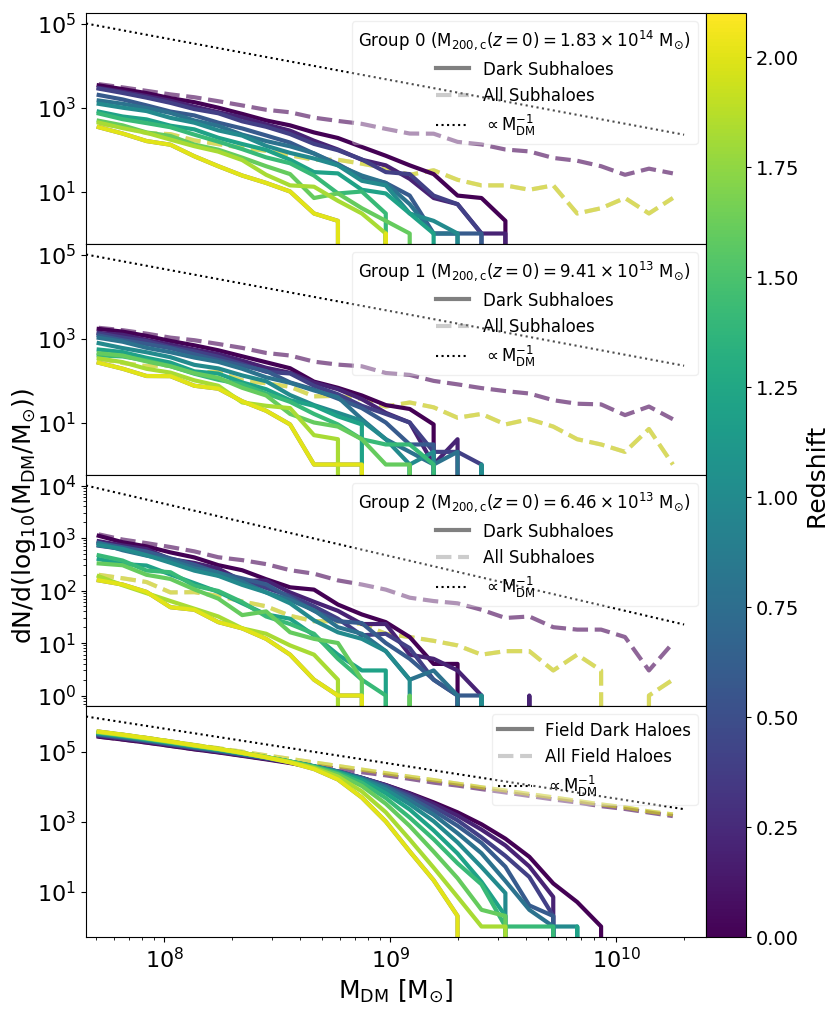}
    \caption{\textit{Top 3 Panels:} Subhalo mass functions for the three galaxy groups and clusters considered in this study over the redshift range $z = 0-2$. We consider only the subhaloes associated to these systems in the \textsc{subfind} catalogs. \textit{Bottom Panel:} Halo mass functions for field haloes, that is, those that are considered their own centrals. In all panels, the mass functions of all (sub)haloes are shown in the dashed lines, and for the entirely dark (sub)haloes ($\rm M_{*} = 0, \ M_{gas} = 0$) in the solid lines. All lines are colored by the redshift at which they are measured. For comparison, we show the expected slope of the halo mass function in the black dotted line \citep[see ][ and references therein]{Bullock2017}.}
    \label{fig:mass_func}
\end{figure}

To begin to characterize the dark objects that reside within the simulation volume, we first consider the mass function of all low mass subhaloes considered in this study.  The top three panels of Fig.~\ref{fig:mass_func} show the subhalo dark matter mass functions for the three galaxy groups and clusters considered in this study (Groups 0, 1, and 2) and their main progenitors at higher redshift, with the bottom panel showing the same but for field haloes within the considered redshift range.

In Fig.~\ref{fig:mass_func}, we can see the expected slopes in the mass functions for subhaloes in both the groups and clusters and haloes in the field (dashed lines colored by redshift). Following \citet[][and references therein]{Bullock2017}, we find that subhaloes in clusters follow a shallower slope than field haloes, which follow $\rm dn/d(log_{10}(M_{DM}) \propto M_{DM}^{-1}$. We do not see a significant flattening out of the slope in the low mass end that would indicate the presence of resolution effects.

Focusing on what is entirely dark ($\rm M_{*} = 0, M_{gas} = 0$), we can see some slight differences between the mass functions of the subhaloes and field haloes. While these are not volume densities of (sub)haloes, we do find lower upper-end mass cutoffs and an earlier deviation from the total mass function in clusters than in the field. This difference may be due to environmental affects of the clusters themselves \citep[such as tidal stripping, see][]{Nagai2005} or as a manifestation of the limitations of \textsc{subfind} \citep[see][for an in-depth discussion.]{vandenBosch2018, ForouharMoreno2025}. We thus iterate that the masses of of entirely dark subhaloes should be taken as a conservative limit.

\begin{figure}
    \centering
    \includegraphics[width = \columnwidth]{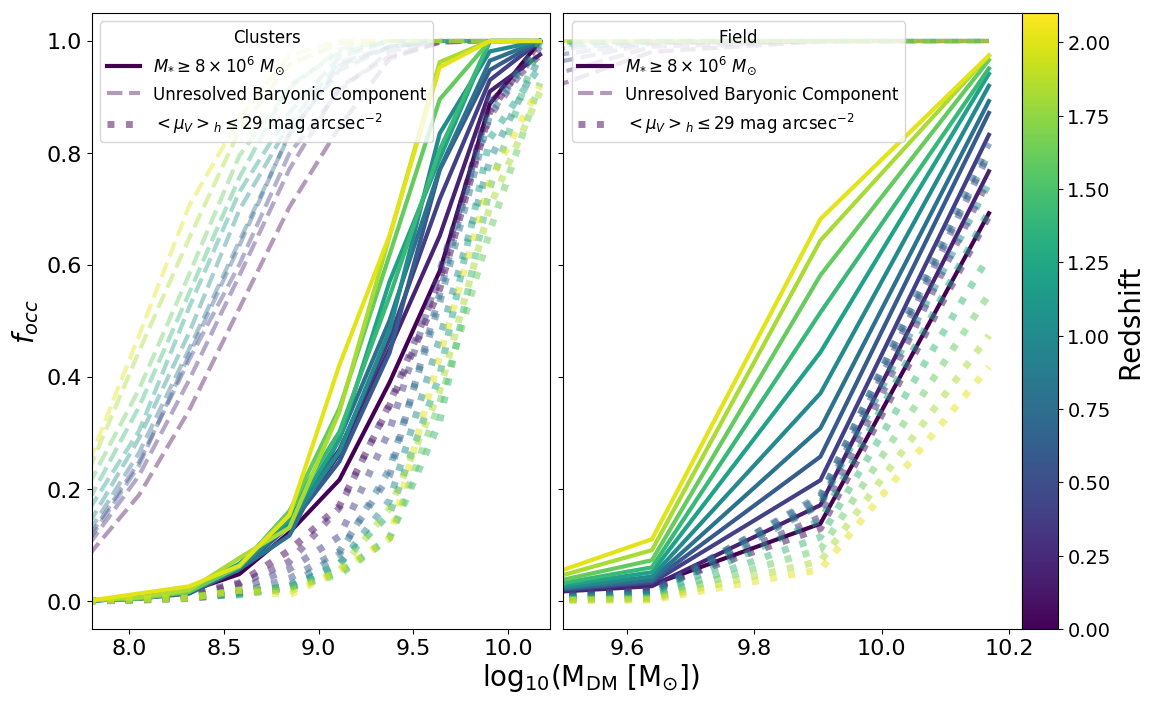}
    \caption{Occupation fraction of dark matter subhalos at different redshifts. We show the occupation fraction of dark matter haloes for three different sets of galaxies: those that are resolved by particle number (solid lines), those that are above a surface brightness limit of $\rm 29 \ mag \ arcsec^{-2}$ (dotted lines), and those that contain at least one star particle (dashed lines). Each curve is colored by redshift. Note that the range over which we show the field occupation fraction is smaller than for subhaloes in clusters. We can see that a small number of subhaloes on the upper end of our probed mass range would be considered observationally dark.}
    \label{fig:occ_frac}
\end{figure}

\subsection{Galaxy Occupation Fraction}
\label{ssec:occ}


One aspect that should be noted in the interpretation of entirely dark subhaloes in TNG50 (or similar sort of simulation) is that the stellar mass of the galaxies that they could contain may be smaller than the mass of the baryonic resolution element within the simulation. It could be argued that these dark objects simply host an \textit{unresolved} stellar component. In order to circumvent this limitation, we instead turn our focus to what is \textit{observationally} dark.
As mentioned in Sec.~\ref{sec:methods}, we consider a galaxy resolved if it has at least 100 star particles, and we place an additional surface brightness cut (as observed at $z = 0$) of $\rm <\mu_V> = 29$ mag arcsec$^{-2}$ to mimic an observational cut.

To illustrate the impact of these cuts, we show in Fig. ~\ref{fig:occ_frac} the occupation fraction ($\rm f_{occ}$) of dark matter subhaloes (left) and haloes (right) as a function of total dark matter mass. We define $\rm f_{occ}$ to be the fraction of (sub)haloes at fixed dark matter mass that host a galaxy. We calculate $\rm f_{occ}$ considering three different cuts on baryonic properties: those with at least 100 star particles or $\rm M_{*} \geq 8\times 10^{6} \ M_{\odot}$ (solid lines colored by redshift), those satisfying the surface brightness requirement (dotted lines colored by redshift), and those that host at least one star particle or gas cell (unresolved baryonic component, dashed lines colored by redshift).

At $z = 0$ (dark purple curves) the surface brightness cut implies a very slightly lower occupation fraction for subhaloes with $\rm M_{DM} \lesssim 10^{9.5} \ M_{\odot}$. In the field, we find that the surface brightness cuts give an occupation fraction this is much less different than what is found in clusters using particle counts, suggesting that these galaxies should be more luminous or, at least, more compact. 
Additionally, the closeness between the occupation fraction inferred from particle counts and surface brightness cuts at $z = 0$ is expected in both the field and in cluster, as the surface brightness cut is on par with relatively deep imaging from Hubble \citep[see e.g., the Frontier field][]{Coe2015, Lotz2017}.

We do see more of a deviation in the occupation fractions measured from particle counts and surface brightness as redshift increases. This is additionally to be expected, as it becomes more observationally challenging to observe low surface brightness galaxies at higher z. This being said, the surface brightness limits presented here are affected by the too-large galaxy sizes predicted by the TNG model at low masses \citep[see][ for a more detailed discussion]{pillepich2018}. The occupation fraction, as measured by surface brightness limits within the simulation, should thus be taken as a conservative upper limit on what is considered observationally resolved.

Investigating $\rm f_{occ}$ including (sub)haloes that possess an unresolved baryonic component (dashed lines in Fig. ~\ref{fig:occ_frac}), we see much lower mass objects hosting a potentially luminous component. This is expected, and serves simply to show the impact of moving the ``luminous'' or resolved threshold in this study. It does, however, show the additional interpretation that many more low mass galaxies may exist than what our surface brightness limit can probe. In some respect, 
we could regard this as a manifestation of something like a missing satellites problem, which recent work shows is no longer in tension with $\Lambda$CDM \citep[e.g.][and references therein]{Sales2022}.

Keeping these limitations in mind, and combining these results with those presented in Fig ~\ref{fig:mass_func}, we find a conservative upper limit of the most massive completely dark subhalo of $\rm M_{DM} \sim 10^{9.5} \ M_{\odot}$ in the considered galaxy groups and clusters. These more massive dark objects are exceedingly rare, and tend to only exist at relatively low redshift ($z \lesssim 0.2$). Within the observationally dark limit ($\rm <\mu_V>_h \gtrsim 29 \ mag \ arcsec^{-2}$) in the same environments, we place a similarly conservative upper limit of $\rm M_{DM} \sim 2\times 10^{10} \ M_{\odot}$, which are predominantly present at $z \gtrsim 1$. 
Extending to the entire TNG50 volume, we find that massive ($\rm M_{DM} \gtrsim 10^{10} \ M_{\odot}$), observationally dark objects are more present at all z. We find the most massive entirely dark field object appears at $z = 0$, and has a mass of $\rm M_{DM} \sim 10^{9.75} \  M_{\odot}$. 

These results may be volume limited. Investigating dark field haloes in the TNG100 simulation \citep[][which features the same galaxy formation model as TNG50, but with $\sim 16 \times$ lower resolution]{nelson2018, pillepich2018}, we find a most massive entirely dark object of $\rm M_{DM} = 3.5 \times 10^{10} \ M_{\odot}$. Some resolution effects are likely to play a role in this difference as well, so higher resolution studies in larger volumes are necessary to disentangle these effects.

Finally, at any redshift, we expect there to be many more subhalos that host a low-mass luminous component than what is observed. 
With better telescopes and deeper surveys, we may yet find that the detection of subhalos thought to be dark actually contain low-luminosity galaxies.For the upcoming Habitable Worlds Observatory, for example, using a setup involving a 6.5 m mirror, it would be possible to observe to even lower surface brightnesses than what is presented in this study. For example, for an object of surface brightnesses of $\rm 30 \ mag arcsec^{-2}$ could be observed in a $\sim 1$ hour exposure with a signal to noise of $\sim 10$ in the V band\footnote{\hyperlink{https://hwo.stsci.edu/camera_etc}{https://hwo.stsci.edu/camera\_etc}}.

\subsection{Distribution of Dark Subhalos in Galaxy Clusters}

\begin{figure*}
    \centering
    \includegraphics[width = \textwidth]{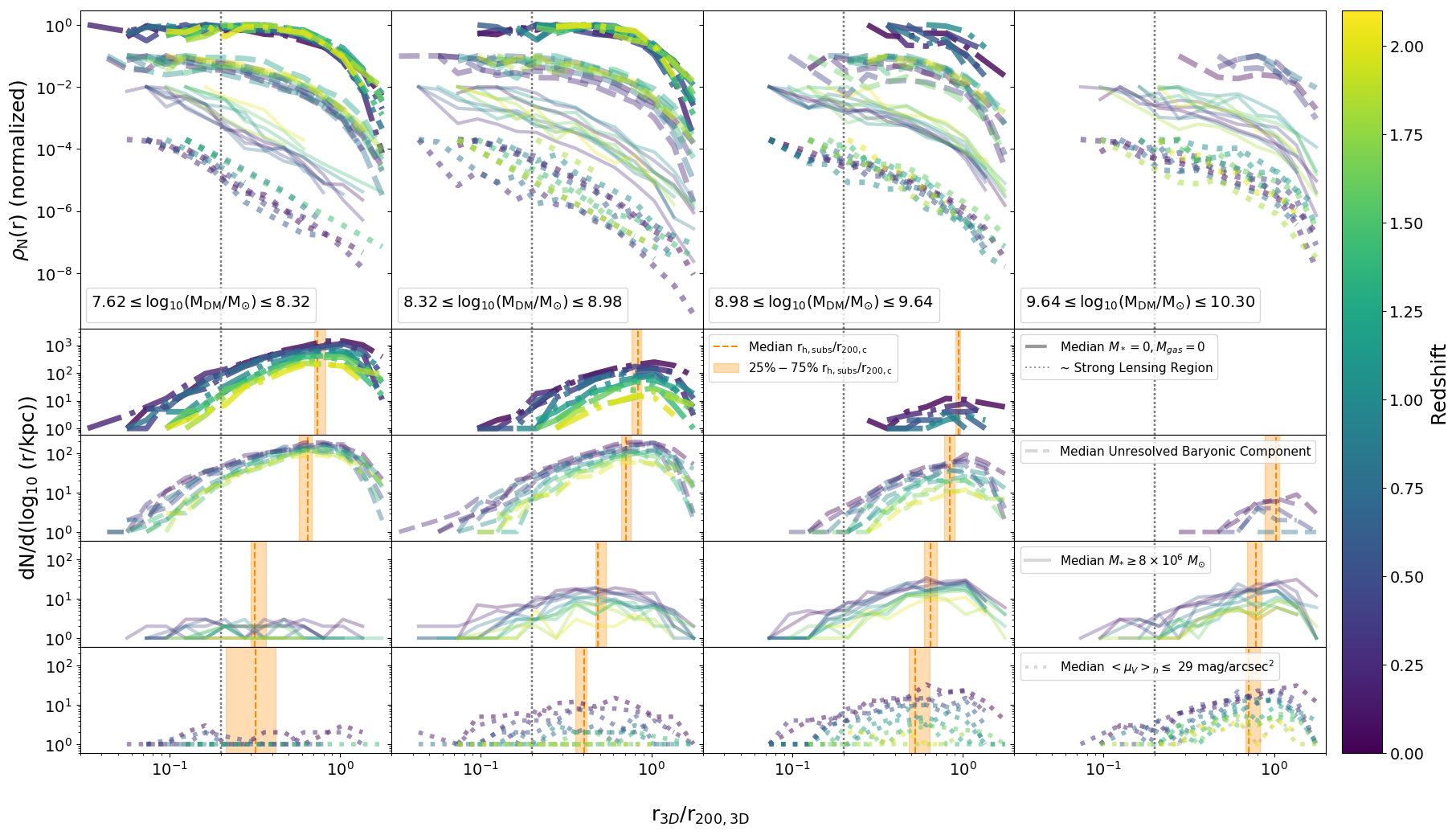}
    \caption{\textit{Top panels:} The number density of subhaloes as a function of $\rm r/r_{200,c}$ in our targeted galaxy groups and clusters over time for subhaloes in the mass range $\rm 7.62 \leq log_{10}(M_{DM}) \leq 8.32$ (leftmost), $\rm 8.32 \leq log_{10}(M_{DM}) \leq 8.98$ (middle left), $\rm 8.98 \leq log_{10}(M_{DM}) \leq 9.64$ (middle right), and $\rm 9.64 \leq log_{10}(M_{DM}) \leq 10.30$ (rightmost panel). \textit{Lower panels:} Number of subhalos per $\rm log_{10}(r/kpc)$, $\rm dN/dlog_{10}(r)$ as a function of $\rm r/r_{200,c}$ for the same mass bins. Each lower panel shows a different component of subhaloes: $\rm M{*} = 0, \ M_{gas} = 0$ (opaque dot dashed lines colored by redshift), those with an unresolved baryonic component (translucent dashed lines), those with $\rm M_{*} \geq 8 \times 10^{6} \ M_{\odot}$ (transparent lines), and those with $\rm <\mu_V>_h \leq 29 \ mag \ arcsec^{-2}$ (translucent dotted lines). Each line in each panel represent the median value of number (density) between the three target cluster at each radial bin at each redshift. For comparison, we show the median half number radii and their 25-75 percentile spreads in the orange dashed line and shaded region respectively and the strong lensing region for the most massive galaxy cluster at $z = 0$ as the gray dotted line. We can see that while they have a greater half-number radius, dark and observationally dark subhaloes of mass $\rm log_{10}(M_{DM}/M_{\odot} \leq 8.98$ are most likely, by number, to reside in the strong lensing region of our target groups and clusters at all redshifts.}
    \label{fig:dist}
\end{figure*}

Looking for dark substructure within lensing clusters requires such objects to be close enough to images to distort them \citep[e.g.,][]{Vegetti2012, Vegetti2014, Despali2024}. Understanding the plausibility of encountering a sufficiently massive substructure along such an advantageous line of sight requires additional understanding of how dark substructures distribute within their host galaxy groups or clusters. While other recent works probe the properties of (dark) subhaloes themselves within TNG50, \citep[e.g.,][]{Despali2024}, this work investigates how (dark) subhaloes distribute themselves in galaxy group and cluster environments.

We characterize the distribution of subhaloes belonging to our target clusters and their progenitors in Fig. ~\ref{fig:dist}. The top panels show the median normalized number density of our target systems' subhaloes as a function of radius at each considered redshift. From left to right, we show this for bins of increasing mass. As in Sect.~\ref{ssec:occ}, we investigate three different sets of objects: galaxies resolved by at least 100 star particles (solid lines), those that exceed an average surface brightness within the stellar halfmass radius of $\rm 29 \ mag \ arcsec^{-2}$ (transparent dotted lines), and those that are assigned at least one star particle (dashed lines). Additionally considered here are subhaloes that are entirely dark ($\rm M_{*} = 0, \ M_{gas} = 0$, dash-dotted lines). All sets of number density profiles are arbitrarily normalized for ease of interpretation.

Of note in Fig.\,\ref{fig:dist} is that we see some consistency in the slope and shape of each set of luminous or dark subhalos as a function of redshift. This is expected given the self-similar nature of $\Lambda$CDM. Indeed, as found in previous works investigating the distribution of subhalos in galaxy clusters \citep[e.g., ][]{Wu2013, Han2016}, that only the normalization of the mass profiles changes between mass bins.

A more stark disagreement with the predictions from the simulations presented in e.g., \citet[][]{Han2016, Nagai2005} is the shape/slope between the two luminous cuts, the observationally dark sample, and the truly dark subhalos. Particularly in the lower mass bins (those between $7.70 \leq \rm log_{10}(\rm M_{DM}/\rm M_{\odot}) \leq 9.42$. Combining the expected NFW-type shape of subhaloes of a given mass within a galaxy cluster with an expected occupation fraction of galaxies in a mass bin, results in a luminous component that is consistent in shape and slope with the profile of the entirely dark subhaloes. While there is some variance in the shape due to the small number statistics in the lowest mass bin, this is insufficient on its own to explain the discrepancy in slope seen in higher mass bins. 

Two other effects are at play in the shaping of the profiles for luminous galaxies ($M_* \geq 8 \times 10^6$ M$_{\odot}$ and those with $<\mu_V>_h \leq 29 ~\rm mag \ arcsec^{-2}$. One is an obvious case of resolution. The other effect, perhaps more physical, comes from a bias in the sampling of the stellar mass - halo mass relation. With the mass and brightness cuts applied in this work, we can only trace luminous galaxies in lower-mass dark matter halos for their stellar mass, biasing for tracing objects that have either experienced significant tidal stripping or that formed later in time. See Appendix ~\ref{app:res} for more details.

The bottom four panels of Fig. ~\ref{fig:dist} show the number of subhaloes in given bins of $\rm log_{10}(r)$ as a function of $r/r_{200,c}$. No normalization has been applied to these curves. Within the inner regions of our selected systems over time, we can see that the subhaloes with $\rm 7.62 \leq log_{10}(M_{DM}/M_{\odot}) \leq 8.98$ are particularly prevalent. The vast majority of these objects are observationally dark in some way in that they either possess some unresolved baryonic component or no baryons at all. In the lowest mass bins, very few resolved ($\rm M_{*} \geq 8 \times 10^6 \ M_{\odot}$) or luminous ($\rm <\mu_V>_h \leq 29 \ mag \ arcsec^{-2}$) subhaloes are present, but they tend to distribute at all radii in the clusters. As discussed in Sec. ~\ref{ssec:mass_func} and ~\ref{ssec:occ}, entirely dark objects with masses $\rm M_{DM} \gtrsim 10^{9.5} \ M_{\odot}$ are exceptionally rare in clusters (and in the field). While there are a handful of dark subhaloes around this dark matter mass in the most massive clusters in the simulation, the median case is that there are no massive dark subhaloes at any radius within the cluster at any redshift. We do emphasize that only one relatively massive cluster exists in TNG50, so as a limiting case, the presence of massive dark subhaloes in cluster environments can be more conservatively denoted exceptionally rare in the TNG model.

More quantitatively, we investigate the half number radius of subhaloes, which we define as the radius that contains half the number of subhaloes in a mass bin within a given group or cluster (denoted by the orange dashed line and the orange shaded region). While we do find that low-mass (observationally) dark subhaloes tend to be more numerous in the centers of groups and clusters, the half number radius of these subhalos tend to be 0.8-0.9 $\rm r_{200,c}$ for dark subhaloes and 0.5-0.8 $\rm r_{200,c}$ for observationally dark objects that host an unresolved baryonic component. For subhaloes with resolved and luminous baryonic components, we find much closer half-number radii, between $\rm \sim 0.2-0.5 \ r_{200,c}$ for the two lowest mass bins, and $\rm \sim 0.5-0.8 \ r_{200,c}$ for higher mass objects. 
This mass segregation is expected and can be explained through the mass-dependent timescale of disruption \citep[e.g., through dynamical friction, see ][]{Nagai2005, bandt2008, Han2016, O'Shea2024}. Additionally, the most massive dark subhaloes found within the groups and clusters themselves are very close in mass to the most massive dark halos in the field. Thus, they may be the result of recent infall.

\section{Discussion: Implications for Dark Subhalo Searches}
\label{ssec:lens}

\begin{figure}
    \centering
    \includegraphics[width=\columnwidth]{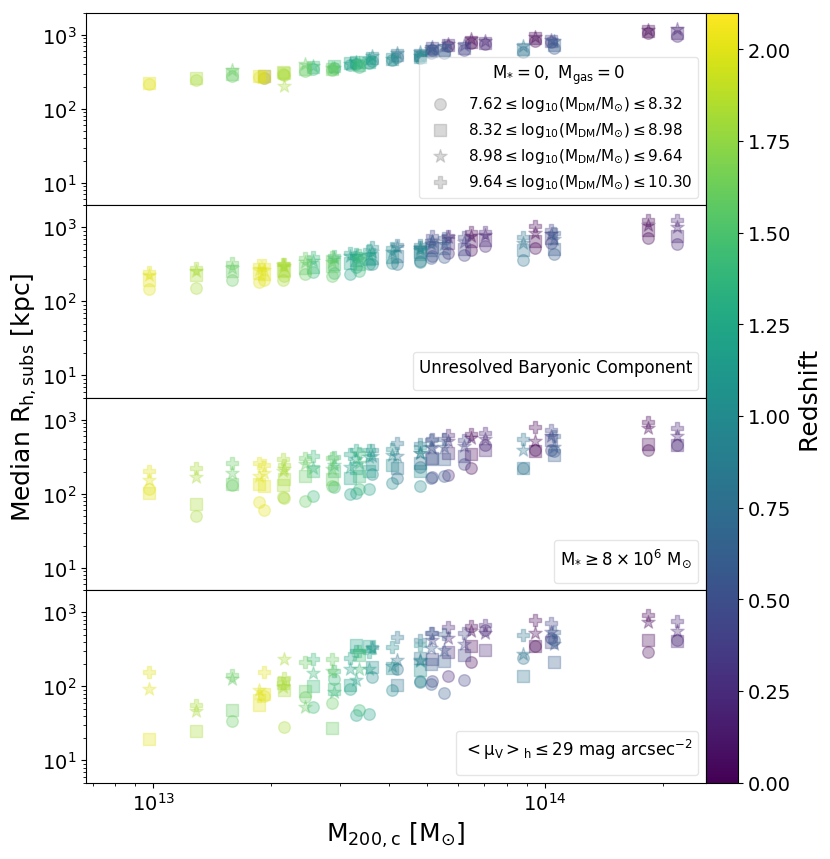}
    \caption{Median half number radius for dark (top), unresolved (panel 2), resolved (panel 3), and luminous (bottom) of 36 projected directions centered on each cluster progenitor. We can see that even in projected space, we are more likely to find lower mass subhaloes (circles and squares) closer to the centers of their hosts than higher mass subhaloes (stars and plus signs). Interestingly, the half number radius of dark things is roughly half the radius of the projected area, suggesting that these objects are roughly uniformly distributed in a projected configuration.}
    \label{fig:proj_rh}
\end{figure}

The previous sections have illustrated that there are several complicating factors in understanding truly dark subhaloes in galaxy groups and clusters. Within the target systems we trace over time in TNG50, and at all redshifts, we find very few massive dark (sub)haloes. This is not unexpected: galaxy formation theory asserts that all haloes with a virial temperature, $\rm T_{vir} \sim 2000 K$, or, equivalently, a mass $\rm M_{200,c} \gtrsim 10^{9.7} \ M_{\odot}$, can cool gas and form stars \citep[see, e.g.,][]{Rees1986, Efstathiou1992, Thoul1996, Quinn1996, Barkana1999, Sawala2016, Benitez-Llambay2020}. This is reflected quite clearly in the occupation fraction of resolved galaxies in field haloes (see Fig.~\ref{fig:occ_frac}). These objects are expected to lose mass after they infall into their present-day group or cluster \citep[e.g.,][]{Nagai2005, bandt2008, Han2016, O'Shea2024}, so massive dark objects are expected to be more rare in clusters. 

Our conclusions have, thus far, been investigated in 3D space rather than in projection. The effects of (dark) objects on a strongly lensed image is cumulative; objects in the field along the line of sight need to be taken into account. As an albeit simplistic test of the impact of line of sight haloes on our previous results, we have taken ``observations'' of our target groups and clusters as though they are centered in a cylinder the length of the simulation volume ($\rm L = 51.7 \ Mpc$) with a radius $\rm 2R_{200,c}$ from their centers along a grid of 36 different sight lines. Fig.~\ref{fig:proj_rh} shows the resulting median projected half-number radius of (sub)haloes their respective mass bins and populations---those that are dark ($\rm M_* = 0, M_{gas} = 0$), those that have an unresolved baryonic component, those that are resolved with $\rm M_* \geq 8\times 10^{6} \ M_{odot}$, and those that are luminous ($\rm <\mu_V>_h \ \leq 29 \ mag \ arcsec^{-2}$)---as a function of the host's halo mass, $\rm M_{200,c}$. 

As in Fig.~\ref{fig:dist}, we see the similar, though expected, trend that lower mass subhaloes (circles and squares colored by redshift of the host group) tend to reside closer to the centers of their hosts than higher mass subhaloes (stars and plus signs). We do, however, though somewhat unsurprisingly, find that the half number radii, when deprojected \citep[assuming $\rm R_{proj} = 0.75 r_{3D}$ as in, e.g.,][]{Hernquist1990, Wolf2010, Somerville2018} are larger for the set of entirely dark objects (top panel of Fig.~\ref{fig:proj_rh}) and those with an unresloved baryonic component (following panel of Fig.~\ref{fig:proj_rh}) than would be expected of the distribution of subhaloes when considering groups and clusters in isolation. Conversely, we see smaller half number radii for objects that host a resolved stellar component ($\rm M_{*} \geq 8 \times 10^6 \  M_{\odot}$, next to last panel of Fig.~\ref{fig:proj_rh}), and those that are luminous ($<\mu_V>_h \leq 29 \rm mag \ arcsec^{-2}$, bottom panel of Fig.~\ref{fig:proj_rh}).

Both of these effects can be traced to the differing occupation fraction of subhaloes in groups and clusters and field haloes. As can be seen from Fig.~\ref{fig:occ_frac}, there is a higher-mass turnover in the occupation fraction for all considered sets of galaxies. Thus, (observationally) dark objects, the increase of line of sight objects causes a flattening out of the radial distribution. This acts to pull the half-number radii further out. For (sub)haloes that satisfy the star particle number cut and surface brightness cut, we find radial profiles that are \textit{steeper} than expected considering only the target cluster in isolation. This is particularly prevalent in the lowest mass bin (circles in Fig.~\ref{fig:proj_rh}), where the radial profiles are already quite steep as a result of our chosen stellar mass and surface brightness cuts.

The detection of a truly dark subhalo could provide test of both $\Lambda$CDM and our current galaxy formation framework. In order for such objects to be (indirectly) observationally detected at galaxy group or cluster scales by invoking strong gravitational lensing, a dark subhalo would need to exist very near the strong lensing region ($\rm R \sim 10-250 \ kpc$ from the center of a cluster). We show a radius of $\rm 250 \ kpc$ for the most massive cluster in TNG50 at $z = 0$ by the vertical black dotted line in Fig.~\ref{fig:dist}. Investigating the number of subhaloes at this radius at $z~0$, we find a total of $~100$ entirely dark subhaloes that mostly reside in the mass range $\rm 7.62 \leq log_{10}(M_{DM}/M_{\odot}) \leq 8.32$, with a similar total number of subhaloes with an unresolved baryonic component, most of which exist in the range $\rm 7.62 \leq log_{10}(M_{DM}/M_{\odot}) \leq 8.98$. There are $\sim20$ total subhaloes that host a resolved or luminous galaxy. Thus, as expected from a $\Lambda$CDM framework, low mass objects, in particular, those that are found to be (observationally) dark in this study, are most likely to be found in this region.

\section{Summary}
\label{sec:summary}

In this study, we have utilized the three most massive galaxy groups and clusters, as well as their progenitors from $\rm 0 \leq z \lesssim 2$, in the TNG50 simulations to investigate their populations of low-mass subhaloes ($\rm 4.54 \times 10^7 \leq M_{DM}/M_{\odot} \leq 2\times 10^{10}$). We find that: 
\begin{itemize}
    \item Massive dark (sub)haloes ($\rm M_{DM} \gtrsim 10^{9.7} \ M_{\odot}$ are rare, which is expected from galaxy formation theory, which places a mass limit of $\rm M_{200,c} \lesssim 10^{10} \ M_{\odot}$ on objects that will always form stars.
    \item While it is unsurprising that faint objects are not easily detected at any redshifts, they are particularly relevant to consider in subhalo studies. With an inferred occupation fraction of $f_{occ} < 1$ still by $\rm M_{DM} = 10^{10} \ M_{\odot}$ at $z \sim 2$, it is entirely feasible that a faint dwarf galaxy may reside in detected massive subhaloes. This discrepancy alleviates itself at lower redshift, simply by nature of an increased ease of observation of low-mass objects.
    \item Both dark and observationally dark subhaloes reside at all radii in galaxy groups and clusters. It is low mass, $\rm M_{DM} \lesssim 10^{8.32} \ M_{\odot}$, subhaloes that are most likely, by number, to be dark and reside in the strong lensing regions of galaxy groups and clusters.
    \item Taking into account the effects of line of sight haloes and subhaloes, we find flatter distributions of dark and observationally dark subhaloes within $\rm 2 R_{200,c}$ of their host groups and clusters at all redshifts. However, we similarly find it rare for dark subhaloes with masses $\rm M_{DM} \gtrsim 10^{9} \ M_{\odot}$ to exist in the strong lensing regions of groups and clusters.
    \end{itemize}

It is important to note that the above results are \textit{conservative} estimates for the expected properties of dark and observationally (sub)haloes. First, we have sacrificed working with larger volumes with higher mass galaxy clusters in favor of increased resolution in this study. There may be volume effects that are outside of the scope of this study to investigate. Second, low mass galaxies in the TNG galaxy formation model tend to be puffier than expected, meaning that the predicted surface brightnesses are likely lower than expected as well. We thus expect an over-prediction of observationally dark objects \citep[see ][]{pillepich2018}. Finally, the resolution limit may imply that the absence of stars and gas in entirely dark objects is rather indicative of simply possessing a stellar or gaseous component that is below the resolution limit of the simulation. Finally, we thus emphasise that our results should be taken as upper limits for both mass and number of dark and observationally dark objects.

Observational tests are, of course, necessary to confirm the predictions set forth in this work. Several studies have made potential detections of subhaloes with masses greater than $\rm 10^{9} \ M_{\odot}$ \citep[e.g.,][]{Vegetti2012, Vegetti2014, Lange2025}. Assuming the conditions explored in this study, it is unlikely that they should be dark. Therefore, to probe the low-mass objects that should be numerous in the strong lensing regions of galaxy clusters, extremely high-resolution imaging and extremely deep observations are necessary, such as that anticipated for the Habitable Worlds Observatory.

\section*{Acknowledgements}

JED, MJ and DJL are supported by the United Kingdom Research and Innovation (UKRI) Future Leaders Fellowship `Using Cosmic Beasts to uncover the Nature of Dark Matter' (grant number MR/X006069/1). DJL is also supported by Science and Technology Facilities Council (STFC) grants ST/T000244/1 and ST/W002612/1. AF acknowledges support of a UKRI Future Leaders Fellowship (grant no. MR/T042362/1) and a Wallenberg Academy Fellowship. This work used the DiRAC@Durham facility managed by the Institute for Computational Cosmology on behalf of the STFC DiRAC HPC Facility (www.dirac.ac.uk). The equipment was funded by BEIS capital funding via STFC capital grants ST/K00042X/1, ST/P002293/1, ST/R002371/1 and ST/S002502/1, Durham University and STFC operations grant ST/R000832/1. DiRAC is part of the National e-Infrastructure.


\section*{Data Availability}

The work presented in this paper utilizes halo catalogs and merger trees from the Illustris-TNG project \citep{nelson2019dr}. The data are publicly available at \href{https://www.tng-project.org/}{https://www.tng-project.org/}.



\bibliographystyle{mnras}
\bibliography{subhalo} 




\appendix



\section{The effect of mass cuts on the radial distribution of subhaloes}
\label{app:res}

\begin{figure}
    \centering
    \includegraphics[width=\columnwidth]{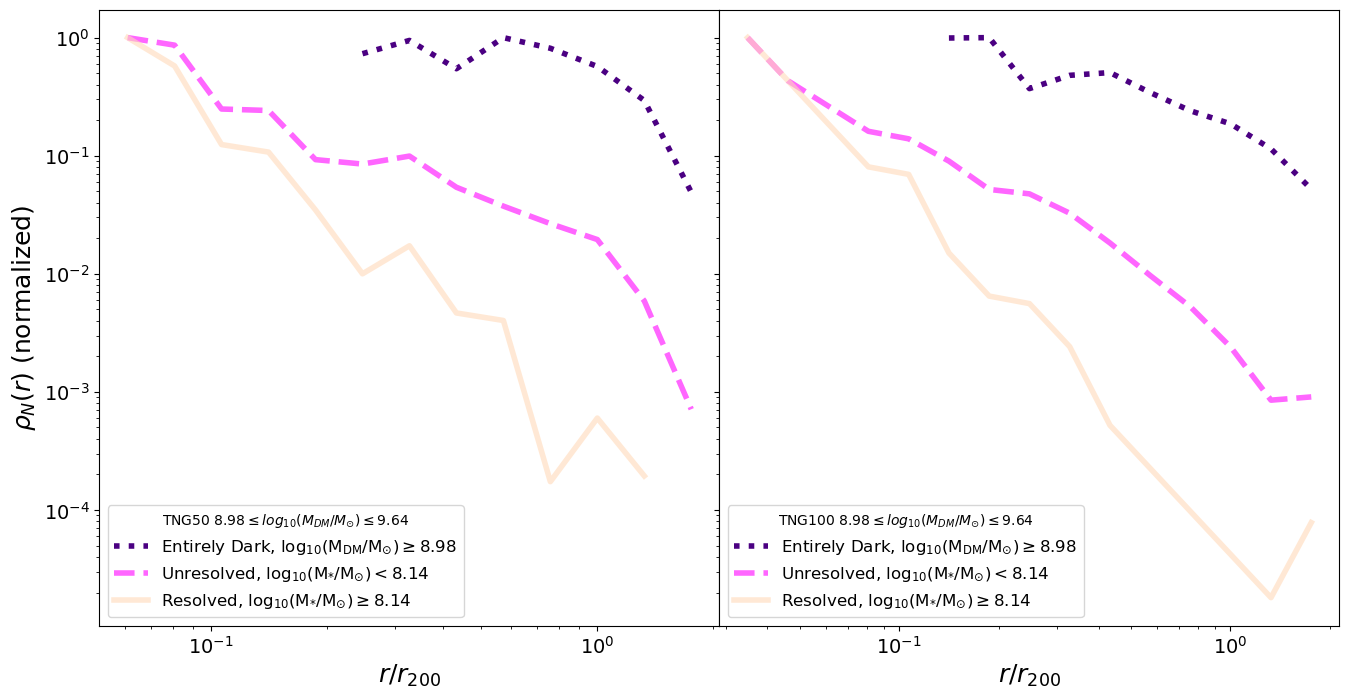}

    \includegraphics[width = \columnwidth]{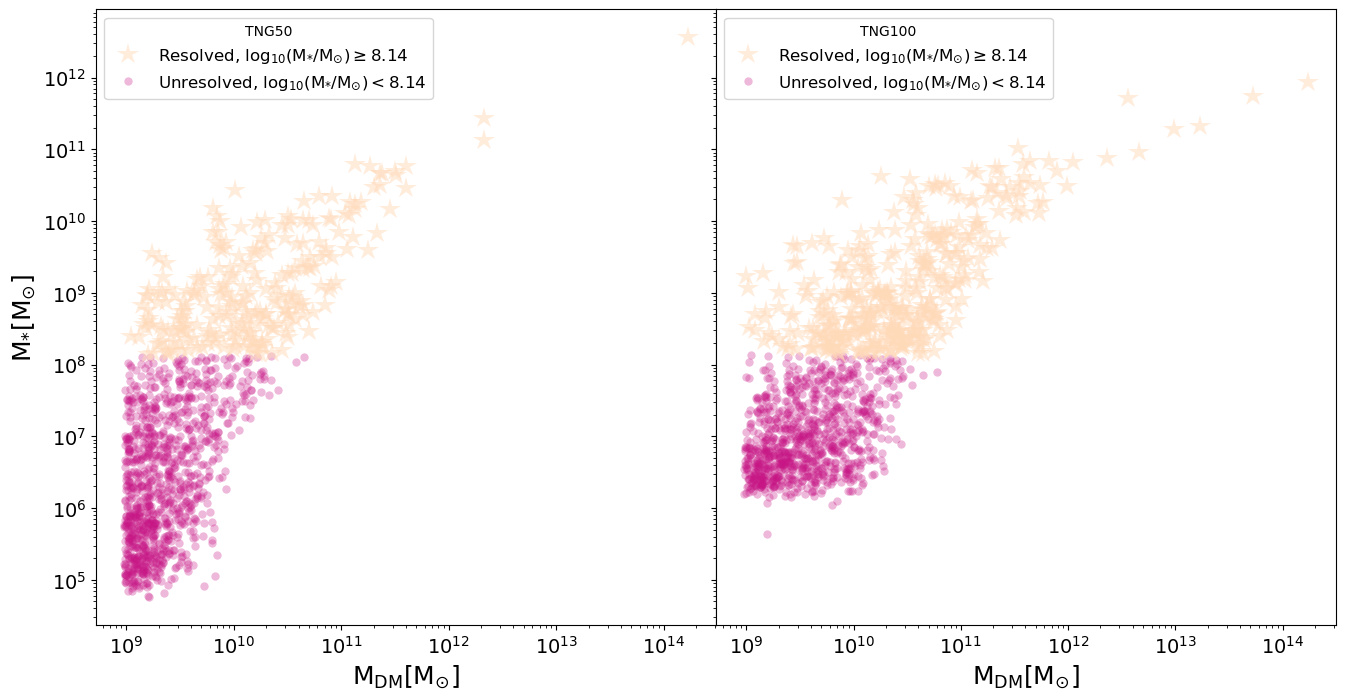}
    
    \caption{\textit{Top:} Radial number density profiles that have been normalized to their maximum value for TNG50 (left) and TNG100 (right) in their lowest overlapping mass bin ($\rm 8.98 \leq log_{10}(M_{DM}/ M_{\odot}) \leq 9.64$). While this mass range is well resolved in TNG50, we see the same behavior as in TNG100, which has a lower resolution and for the lowest mass bin of TNG50. \textit{Bottom}: stellar mass vs dark matter mass for TNG50 (left) and TNG100 (right). The behavior in the radial number density profiles appears to come from the cutoff in the stellar mass-dark matter mass relationship, which does not affect higher mass bins. We conclude that this effect is not an issue of numerical resolution so much as it is an affect of our choices of mass cuts. Interestingly, the lack of entirely dark subhaloes at small radii occurs even in the lower resolution scenario.}
    \label{fig:enter-label}
\end{figure}

To better explain the difference in slope seen in our lowest mass bins in Fig. \ref{fig:dist}, we investigated the impacts of small number sampling, resolution, and  mass/brightness cuts, 

We can see in the two lowest mass bins (two left most panels) in Fig. \ref{fig:dist} that there are very few subhalos that satisfy requirements of being luminous. We thus took the distrbution of all subhaloes in these mass bins, regardless of stellar content, and subsampled a fraction equal to the average occupation fraction in the bin. This does introduce effects of discreteness into the resulting radial profiles, but they roughly maintain the shape of the underlying profile of subhaloes. This is to be expected, as this should effectively change the normalization of the radial profile. But this condition is insufficient to recreate the behavior we find in the figure.

We thus turned to the highest resolution run of TNG100 to investigate the impacts of resolution. Using only groups within the mass range considered in this study, we found that in the lowest mass bin that we could resolve using TNG100 (roughly equivalent with the third panel in Fig \ref{fig:dist}), imposing a particle number cut of 100 stars reproduced the power-law behavior. 

As a sanity check, placing the same mass constraints from TNG100 on the TNG50 data returned the same result in that mass bin. Given that these objects are well resolved in this study, we found that mass cuts are equivalent to introducing a sampling bias: we only probe part of the stellar mass-halo mass relation when we introduce a mass cut. This biases our search only to galaxies that reside in less massive DM halos for their stellar mass. Placing these cuts anywhere with a significantly low occupation fraction returns these results. We therefore consider the difference in slope is only in part caused by resolution, and is more so an affect on our chosen mass cuts.



\bsp	
\label{lastpage}
\end{document}